%% file: blindsignedid.tex
\documentclass[letterpaper,twocolumn,10pt]{article}

\usepackage{xspace,amsmath,graphicx,subcaption}
\usepackage[font=small,labelfont=bf]{caption}
\usepackage{amssymb,amsthm}
\usepackage{authblk}
\usepackage{color}
\usepackage{hyperref}
\usepackage{titlesec}
\usepackage{marvosym}

\usepackage{url}

\usepackage{breakurl}

\newcommand{\sysid}{\textsf{\mbox{BlindSignedID}}\xspace}
\newcommand{\sysids}{\textsf{\mbox{BlindSignedIDs}}\xspace}

\newcommand{\vote}{\textsf{\mbox{FinalTrial}}\xspace}

\newcommand{\ephid}{\textit{EphID}\xspace}
\newcommand{\ephids}{\textit{EphIDs}\xspace}
\newcommand{\sk}{\textit{SK}\xspace}

\newcommand{\prg}{\textit{PRG}\,\xspace}
\newcommand{\prf}{\textit{PRF}\,\xspace}
\newcommand{\sign}{\textit{SD}\xspace}
\newcommand{\auth}{\textit{Auth}\xspace}
\newcommand{\signpos}{\textit{SP}\xspace}

\newcommand{\modn}{\:\mbox{mod N}\xspace}
\newcommand{\cncat}{\:||\:}
\newcommand{\prefix}{\:\mbox{Prefix}\xspace}

\newcommand{\dpppt}{\mbox{DP-3T}\xspace}
\newcommand{\bt}{\textrm{Bluetooth}\xspace}
\newcommand{\ephrl}{\textrm{ephemeral}\xspace}
\newcommand{\inplace}{\emph{in-place}\xspace}
\newcommand{\authtag}{\textrm{authenticator}\xspace}
\newcommand{\authtags}{\textrm{authenticators}\xspace}

\newcommand{\dosatk}{\textbf{DoS attacks}\xspace}
\newcommand{\idspoofatk}{\textbf{identifier-spoofing attacks}\xspace}
\newcommand{\verephids}{\textit{verifiable ephemeral identifiers}\xspace}

\newcommand{\ie}{\emph{i.e.,}\xspace}
\newcommand{\eg}{\emph{e.g.,}\xspace}
\newcommand{\first}{\textsf{(i)}\xspace}
\newcommand{\second}{\textsf{(ii)}\xspace}
\newcommand{\third}{\textsf{(iii)}\xspace}

\newcommand{\paragraphb}[1]{\vspace{0.01in}\noindent{\bf #1}}

\newcommand{\keywords}[1]{\small\textbf{Keywords:} #1}

\newcommand{\at}{\noindent{\small\textnormal{\raisebox{0.3ex}{\MVAt}}}}

\titlespacing*{\section}{0pt}{3pt}{3pt}
\titlespacing*{\subsection}{0pt}{3pt}{3pt}
\begin{document}
    \title{\bf \sysid: Mitigating Denial-of-Service Attacks \\on Digital Contact Tracing}
    \date{\vspace{-5ex}}

    \author{Bo-Rong Chen}
    \author{Yih-Chun Hu}
    \affil{University of Illinois at Urbana-Champaign\\\{borongc2, yihchun\}\at illinois.edu}

    \maketitle

    \input{abstract}


    \input{introduction}
    \input{assumptions}
    \input{statement}
    \input{protocol}
    \input{privacy}

    \input{security}
    \input{evaluation}
    \input{related}
    \input{conclusion}
    \input{acknowledgment}


\end{document}

%% file: abstract.tex
\begin{abstract}
Due to the recent outbreak of COVID-19,
many governments suspended outdoor activities
and imposed social distancing policies
to prevent the transmission of SARS-CoV-2.
These measures have had severe impact
on the economy and peoples' daily lives.
An alternative to widespread lockdowns
is effective contact tracing
during an outbreak's early stage.
However, mathematical 
models (\eg~\cite{ferretti2020quantifying})
suggest that epidemic control 
for SARS-CoV-2 transmission with manual contact tracing 
is implausible.
To reduce the effort of contact tracing,
many digital contact tracing projects 
(\eg PEPP-PT~\cite{pepppt2020}, 
\dpppt~\cite{troncoso2020decentralized}, TCN~\cite{tcn2020},
BlueTrace~\cite{bay2020bluetrace},
Google/Apple Exposure Notification~\cite{google2020apple},
and East/West Coast PACT~\cite{mit2020pact,chan2020pact}) are 
being developed to supplement manual contact tracing. 
However, digital contact tracing has drawn scrutiny from privacy advocates,
since governments or other parties may attempt
to use contact tracing protocols for mass surveillance.
As a result, many digital contact tracing projects
build privacy-preserving mechanisms to limit
the amount of privacy-sensitive information leaked by the protocol.
In this paper, we examine how these architectures
resist certain classes of attacks, specifically \dosatk,
and present \sysids, a privacy-preserving 
digital contact tracing mechanism, which are \verephids
to limit the effectiveness of MAC-compliant \dosatk.
In our evaluations, we showed \sysid can effectively deny bogus
\ephids, mitigating \dosatk on the local storage beyond 90\%
of stored \ephids. 
Our example \dosatk showed that using 4~attackers can 
cause the gigabyte level \dosatk within normal working hours and days.
\end{abstract}

%% file: introduction.tex
\section{Introduction}\label{sec:intro}
In light of the success of several countries'
SARS-CoV-2 containment strategies
based on aggressive contact tracing,
several researchers have developed approaches to
digital contact tracing~\cite{google2020apple,mit2020pact,
pepppt2020,tcn2020,bay2020bluetrace,
chan2020pact,troncoso2020decentralized}.
Because of the privacy concerns inherent
with disclosing user location traces,
several proposals include
mechanisms for privacy-preserving proximity tracing.
These mecachisms make use of a limited-time identity
to reduce an adversary's ability to track
that device from broadcast to broadcast;
these identities are
variously called \ephrl ID~\cite{troncoso2020decentralized},
Temporary ID~\cite{bay2020bluetrace}, and
Rolling Proximity ID~\cite{google2020apple};
in this paper, we will use the term \emph{\ephrl ID} or \ephid.
To reduce the amount of bandwidth needed
to disseminate positive identities,
these \ephrl IDs are generated from a master secret
in a one-way manner, such that the master secret
is sufficient to derive all relevant \ephrl IDs,
but that several \ephrl IDs leak no information
about other associated \ephrl IDs.

In existing designs, a user receiving an \ephrl ID
can neither associate that \ephrl ID
with other \ephrl IDs from the same contact,
nor can the user verify the validity of an \ephrl ID
that it receives;
as a result, a user must store all of the \ephrl IDs
that it encounters across the maximum incubation period
(widely set at 14 days for SARS-CoV-2).
Furthermore, storing these \ephrl IDs in the cloud
faces significant privacy challenges;
if a large fraction of users trust a single cloud storage provider
to store unencrypted \ephrl IDs,
that cloud storage provider would learn extensive privacy-compromising
information about a user's social structure.
As a result, existing schemes~\cite{google2020apple,
mit2020pact,tcn2020,chan2020pact,troncoso2020decentralized}
largely assume that \ephrl IDs
are stored on the user's own device until the incubation period has expired.
However, because of the limited storage available
on commonly-used mobile devices,
an attacker can send a large number of bogus \ephrl IDs,
overwhelming the storage of all but high-end devices.
We call this attack the \dosatk.
In fact, the existing \ephid-based 
designs~\cite{google2020apple,mit2020pact,
pepppt2020,tcn2020,bay2020bluetrace,
chan2020pact,troncoso2020decentralized} 
are widely adapted by many 
countries~\cite{model2020us,model2020jp,
pepppt2020,model2020uk,
bay2020bluetrace}, 
which occupy approximately 12.5\%~\cite{population2020wiki} 
of the world population, thus making these people's 
mobile devices vulnerable
to \dosatk.
In this paper, we propose a design for \sysids,
where receivers can verify \ephrl IDs \inplace,
and senders can generate only a limited number of \ephrl IDs
based on a resource constraint, such as personal phone numbers, 
that must be verified
before privacy-preserving \verephids are issued,
while still providing privacy from the certificate authority
through the use of blind signatures.
The rest of this paper is organized as follows:
in Section~\ref{sec:assumptions}, we make 
system assumptions and describe the attacker model.
In Section~\ref{sec:statement}, we describe the security
issues regarding current designs of digital contact
tracing. Section~\ref{sec:protocol} introduces our 
approach in detail. We provide an analysis of
\sysid in Section~\ref{sec:privacy} and~\ref{sec:security}.
Section~\ref{sec:evaluation} demonstrates our example
\dosatk and \sysid can effectively reduce \dosatk. 
Section~\ref{sec:related} reviews current projects.
We make conclusions in Section~\ref{sec:conclusion}. 

%% file: assumptions.tex
\section{System Assumptions and Threat Model}\label{sec:assumptions}

\subsection{System Assumptions}
Each received \ephid is kept in the user's
local storage for at least 14~days. The stored
\ephid includes additional information (\eg
encountering time and duration), which consumes
fixed-byte spaces in the local storage. Moreover,
the user stores every received \ephid regardless
of any encountering duration for recording and 
checking exposure. Finally, the mobile devices
support \bt Random Private Address~\cite{ble2020}, where
each received \ephid cannot be linked
to the real \bt device address.

\subsection{Threat Model}
The attacker is able to modify the BLE protocol 
such that her device can advertise BLE beacon 
packets with the minimum interval. Namely, 
the transmission interval is not bounded by the
underlaying BLE stack (\eg maximum and minimum
intervals in BlueZ~\cite{bluez2020}).
She specifies the standard BLE flags with arbitrary 
content (\ie spoofing \ephid) such that
other users will receive and keep the packets in the
local storage. Morever, the attacker uses a powerful BLE antenna
for larger transmission ranges.

%% file: statement.tex
\section{Problem Environment and Statement}\label{sec:statement}

\subsection{Existing Contact Tracing Design}

In this section, we describe 
\dpppt~\cite{troncoso2020decentralized} at a high level,
to set the context for our attacks and our approaches.
\dpppt generates \ephrl IDs using:
\begin{equation}\label{eq:ephid}
\begin{aligned}
&\ephid_1 \cncat \dotsc \cncat \ephid_n =\\ 
&\prg ( \prf (\sk_t, \mbox{"broadcast\ key"}) )
\end{aligned}
\end{equation}
where $\ephid_i$ is a 16-byte \ephrl ID, 
$\sk_t$ is secret day seed, $\prf$ is a 
pseudo-random function, and $\prg$ is a pseudo-random 
generator.
To further reduce the amount of information disseminated
for a positive report,
\dpppt generates the next secret day seed
$\sk_t$ by hashing the previous secret day seed $\sk_{t-1}$
with a one-way hash function,
so that the dissemination of one secret day seed
allows contact tracing for all subsequent days.

Each user
generates 288~EphIDs per day, randomly reorders them,
and uses \bt Low Energy (BLE)
to broadcast each EphID for one epoch (5 minutes).
A device receiving such a beacon stores the \ephid,
signal strength information (such as RSSI),
and the date on which the beacon was received,
for a total of 36~bytes per \ephid
for the low-cost design
and 52~bytes for the unlinkable design~\cite{troncoso2020decentralized}.
A receiving device must store these records for 14~days.
When a user tests positive,
her secret day seeds from the 
infectious period will be uploaded to the backend, 
so that other users can fetch seeds and compare 
with their local records to determine whether they have 
been in contact with this particular user.

\subsection{Problem Statement}

Although storing in local space with user-generated 
\ephrl IDs reveals no information from the user, 
it suffers from \dosatk.
Specifically, 
because a user cannot link messages 
from different secret-day seeds (and even if they could, 
the attacker can choose arbitrary secret-day seeds), 
an attacker with a high-power \bt antenna 
can stream newly-generated \ephrl IDs limited 
only by bandwidth. BLE 
has bandwidth of 1--2~Mbps,
so any receiver within transmission range of an attacker
will need to store (based on the overhead specified in the \dpppt paper)
1--2~GB per hour or 8--16~GB for an 8-hour day.
Even when considering the possibility of compression or elimination
of certain data, the incompressible \ephid alone will amount to
between 450~MB and 900~MB per hour,
or 3.6~GB  to 7.2~GB across an 8-hour working day.
Since many phones do not have several gigabytes of available storage, 
such an attacker can cause the victim’s phone to 
run out of space. We believe that the lack of authentication
of \ephrl IDs makes many privacy-preserving contact tracing
approaches~\cite{google2020apple,mit2020pact,
pepppt2020,tcn2020,bay2020bluetrace,
chan2020pact,troncoso2020decentralized} vulnerable to \dosatk.

\subsection{Out-of-Scope Problems}

Our approaches are designed for the attacks listed above,
and not for other, similar attacks. In this section,
we describe similar attacks that we do not attempt to address,
and discuss the design choices that led us to draw these boundaries.

Our \dosatk aims at MAC-and-PHY-compliant attackers.
While previous work has examined jamming attacks against the physical 
layer~\cite{chiang2010cross,desmedt2001broadcast,
kiesler1942secret,popper2010anti}
and other attacks against the MAC 
layer~\cite{awerbuch2008jamming,bellardo2003802,
cardenas2007performance,chang2017securemac,
chang2015simplemac,gupta2002denial,
kyasanur2005selfish},
the aim of digital contact tracing is to be immediately available
on commodity hardware, which includes commodity wireless chipsets.
As a result, mechanisms that require new approaches to the physical-layer
or MAC-layer are non-starters for these protocols.
Furthermore, the level of complexity required to build
jamming devices or MAC-layer attacks
makes these attacks less practical than ones
that can be built on any \bt compatible system.
When future devices may include support for more robust
physical- and MAC-layers,
our work can integrate with these approaches for greater robustness.

%% file: protocol.tex
\section{Proposed Approach}\label{sec:protocol}

In this section, we describe our approach
to the attacks discussed earlier.
First, we examine how \ephrl IDs can be \emph{verified}.
Previous work~\cite{bajaj2012spammers,jakobsson1999proofs,
liu2017torpolice,parno2007portcullis,
von2008recaptcha}
have developed schemes that use
attacker's resource limitations
(\eg ability to compute computational puzzles or solve CAPTCHAs) 
to create fairness between normal users and automated attackers.
Our approach to \sysids builds on this work
to limit the ability of adversaries to create unbounded numbers
of valid \ephrl IDs.
In particular, verifying a user's real identity or phone number
can serve as the basis for generating a set of valid identifiers
in a privacy-preserving way.

\subsection{Generating \verephids}

\paragraphb{\sysids.}
We build \verephids on subliminal-free blind signatures~\cite{chaum1983blind},
illustrating the approach using RSA blind signatures.
The signer first generates a public-private key pair for each day;
we call ($e_t$, $d_t$) the verification and signing exponent
used on day $t-2$ to create credentials for day $t$.
(The modulus $n$ should also change from day to day.)
Both the verification exponent, the modulus,
and a $\prefix_{t}$ (used to pad the \ephid to a length suitable for signing)
are published
in advance of day $t-2$.
On day $t-2$, the user creates a main secret day seed $sk_t$,
and uses a pseudo-random generator on $sk_t$ to generate
$M$ secondary secret day seeds:
\begin{equation}
\sk_{t_1} \cncat \dotsc \cncat \sk_{t_M} = \prg (\prf (sk_t))
\end{equation}
(where $M$ is chosen to reduce the chance
that an attacker generates \ephids corresponding
to a secret key; we suggest $M=100$)
and generates $M$ sets of \ephids using
\begin{equation}
\begin{aligned}
&\{\ephid_1 \cncat \dotsc \cncat \ephid_n\}_M =\\
&\prg ( \prf (\sk_{t_M}, \mbox{"broadcast\ key\ M"}) )
\end{aligned}
\end{equation}

\begin{figure}[t]
	\centering
	\mbox{
		\includegraphics[scale=0.45]{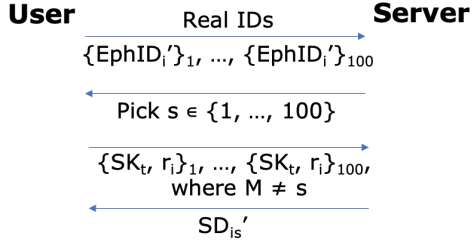}  
	}
	\caption{User Registration.} \label{fig:ephid}
\end{figure}

The user then blinds each \ephid.
To blind \ephids in a manner easily verifiable by the signer,
the user chooses a main blinding seed $b_t$
and uses a pseudo-random generator on $b_t$ to generate
$M$ secondary blinding seeds $b_{t_i}$,
one for each secondary secret day seed:
\begin{equation}
b_{t_1} \cncat \dotsc \cncat b_{t_M} = \prg (\prf (b_t))
\end{equation}
The user then uses a pseudo-random generator on each $b_{t_i}$
to compute a set of blinding values for each \ephid, with
\begin{equation}
\begin{aligned}
\{\hat{r_1} \cncat \dotsc \cncat \hat{r_n}\}_i &= \prg ( \prf ( b_{t_i} ) )\\
r_{i_j} &= \hat{r_{i_j}}^{e_t}
\end{aligned}
\end{equation}
where $e_t$ is the signer's public exponent,
and blinding the \ephid
\begin{equation}
\ephid_{i_j}' := (\prefix_{t} \cncat \ephid_{i_j}) * r_{i_j}
\end{equation}
$\prefix_{t}$ is a value chosen for that day
to pad the \ephid to the size of the RSA group.
For example, if we use 2048-bit RSA
and 104-bit \ephid, the $\prefix_{t}$ is 1944~bits.
In this manner, the user generates $M$ sets of $n$ blinded \ephids,
each of which corresponds to a single secondary secret day seed.
The user then contacts the backend server,
proves her identity
(for example, a phone number, social media account,
or national identity),
and sends the set of all $Mn$ blinded \ephids to the signer.
The signer first verifies the identity
and ensures that it has not previously signed certificates
for this identity for this day.
The signer then
chooses one set of blinded \ephids
asks the user to reveal the blinding factors and
secondary secret day seeds of all remaining sets.
The user then reveals the requested
secret day seeds and their corresponding
blinding seeds.
If all revealed \ephids generated correctly,
the signer blindly signs all resulting values from the 
selected set and sends the signed values back to the user,
as shown in Equation~\ref{eq:sign}. 
\begin{equation}
\sign_{i_s}' := \ephid_{i_s}'^{d_t}\label{eq:sign}
\end{equation}
where $d_t$ is the private key for signing on day $t-2$
(and use on day $t$).
If \ephids
are not correctly generated, the signer
adds the user into a blocklist,
such that the credentials she used are not valid at the signer
for a certain period of time
(the length of time to be chosen should depend on
the possibility of reassignment;
for example, longer-term blocking for national identities
and social media accounts may be reasonable,
but blocks for phone numbers should expire after a few months).
By ensuring that \ephids are generated correctly,
we can ensure that a positive test can be represented by a single
secondary secret-day seed,
ensuring that anyone broadcasting \verephids can correctly report
a positive test.
Figure~\ref{fig:ephid} illustrates
the registration process.

The user computes signatures $\sign_{i}$ from the signed 
values $\sign_{i_j}'$ by inverting the $\hat{r_{i_j}}$
used to blind each $\ephid_{i_s}$:
\begin{equation}
\begin{aligned}
\sign_{i_s} &:= \sign_i'^{d_t} * \hat{r}_{i_s}^{-1} \modn\\
&= ((\prefix_{t} \cncat \ephid_{i_s}) * r_{i_s})^{d_t} * \hat{r_{i_s}}^{-1} \modn\\
&= (\prefix_{t} \cncat \ephid_{i_s})^{d_t} * \hat{r_{i_s}}^{e_t*d_t} * \hat{r_{i_s}}^{-1} \modn\\
&= (\prefix_{t} \cncat \ephid_{i_s})^{d_t} * \hat{r_{i_s}} * \hat{r_{i_s}}^{-1} \modn\\
&= (\prefix_{t} \cncat \ephid_{i_s})^{d_t} \modn\\
\end{aligned}
\end{equation}
All \ephids generated and signed on day $t-2$ are used on day $t$.
We can choose an arbitrary period for signing and refreshing keys,
but we choose a day in line with \dpppt~\cite{troncoso2020decentralized}.

\subsection{Authenticating \verephids in Beacons}

\begin{figure}[t]
	\centering
	\mbox{
		\includegraphics[scale=0.45]{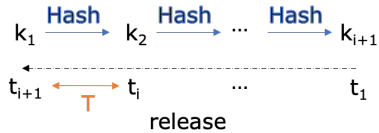}  
	}
	\caption{Key Sequence.} \label{fig:hash_chain}
\end{figure}

Because BLE beacons only have 31~bytes of space,
of which 3~bytes are used for BLE flags,
including a blind signature for an \ephid into the beacon
would require the use of several beacons,
which could lead to computational attacks on packet defragmentation.
To avoid these possibilities,
we propose to have the signer exchange signed \ephids
for \ephids authenticated using TESLA~\cite{perrig2002tesla}.
TESLA uses \authtags based on symmetric cryptography
to provide broadcast authentication
by using the time at which information is disclosed
as the mechanism for creating asymmetry.
In particular, the entity wishing to authenticate messages
(in this case, the signer)
creates a sequence of keys using a one-way hash function:
$k_1, k_2=H(k_1), \ldots, k_{i+1}=H(k_i)\ldots$,
and a schedule at which these keys will be released:
$t_1, t_2=t_1+T, \ldots, t_{i+1}=t_i+T\ldots$, as illustrated
in Figure~\ref{fig:hash_chain}.
The signer will release key $k_i$ at time $t_i$,
so if a verifier receives a message before
the signer's clock has reached $t_i$,
the signer knows that the message is authentic
if it is authenticated using the key $k_i$.
TESLA relies on loose time-synchronization for security,
meaning that each clock must be within a bounded margin of error $\Delta$
from the signer's clock
(in our design, $T$ will be approximately 5~minutes,
and the maximum time-synchronization error will be around 10~seconds).
Thus, if the message is received before time $t_i-\Delta$
on the receiver's clock,
the receiver knows that it is before time $t_i$ on the signer's clock.

\begin{figure}[t]
	\centering
	\mbox{
		\includegraphics[width=\linewidth]{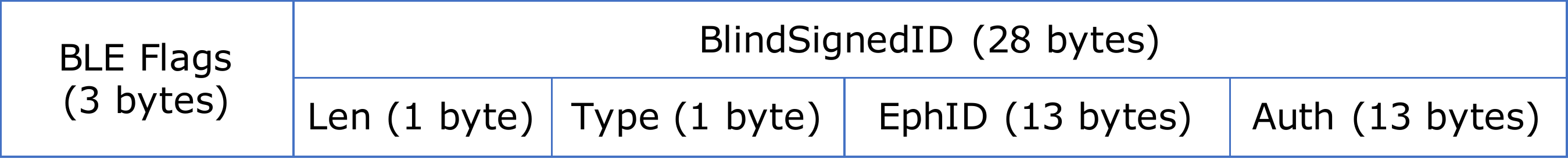}  
	}
	\caption{BLE Beacon Format.} \label{fig:pkt_format}
\end{figure}

To exchange RSA blind-signed \verephids for TESLA-authenticated MACs
corresponding to each \ephid,
a user requests them through a MIX~\cite{chaum1985untraceable}
on day $t-1$,
providing a nonce, an encryption key, the \ephid, 
the corresponding signature \sign, and the time interval
at which the user intends to use them on day $t$
(these time intervals should be standardized across all requesters
to minimize privacy loss).
The signer verifies the signature,
and if valid and not requested before, generates 
the \authtag (\auth) for the requested time interval.
The signer then encrypts the \authtag using the encryption key
and publishes the nonce and encrypted \authtag.
On day $t-1$, a user can retrieve their \authtags 
for day $t$ using one of three approaches: \first full download:
the user downloads the entire list of nonces and encrypted \authtags,
discarding all values other than the ones chosen by the user;
\second partial download:
the user intially constructs the nonces so that they are equal in some bits;
the user then requests a subset of the list by specifying the bits
that are equal across all of that users' requests;
\third individual download: 
a user with limited resources can also 
request their \authtags through anonymous connection 
such as Tor~\cite{tor2020}.
Because \authtags for each \ephid are requested separately through a MIX,
the signer cannot link two \ephids except to know that
\ephids authenticated for the same time
are likely to be from different users.

When broadcasting a beacon, a user broadcasts the \ephid
and the TESLA \authtag used for the current time interval.
When receiving a beacon, a receiver carries the \ephid, the \auth,
the time of receipt, and the signal strength.
A beacon is kept for two time intervals;
if the two keys following the time of receipt
do not match the \auth, then the receiver knows that the \ephid
is not authenticated for the interval in which it was used
(see evaluations in Section~\ref{sec:eval_dos}).
Finally, we use 13~bytes (104~bits) for both \ephids and \auth
to fit within the 28-byte payload limit, 
as shown in Figure~\ref{fig:pkt_format}.

\subsection{Storing Received Beacons}

When a device receives a beacon, 
the user stores them as
$(\ephid_i, \auth_i, \mbox{time}, \mbox{RSSI})$.
When the user obtains the current released 
key $k_i$ at time $t_i$ from
TESLA server, she firstly verifies $k_i$
with $k_{i-1}$ by a one-way hash function;
she then verifies each \ephids
received in the previous period by checking its \authtag $\auth_i$.
If the $\auth_i$ is valid, the device
accepts the record and keeps that record in longer-term storage;
\authtag verification can be performed \inplace 
to mitigate \dosatk.
Optionally, our protocol can be 
combined with~\cite{pietrzak2020delayed} to 
prevent relay and replay attacks. However, 
the details of preventing relay and replay attacks 
are beyond the scope of our protocol.

\subsection{Checking Exposure}

When a health authority determines a user is a positive case, 
all her secondary secret-day seeds and the selected numbers $s$
during the infectious period 
(typically 2 days before symptoms happen) are published
in a blockchain, together with a sequential case number
for the day of publication.
Each user fetches the list of these positive cases,
and generates \ephids with the secondary secret-day seeds 
and the selected numbers $s$, and
compares the \ephids with the \ephids in their local storage.
If there is a match, the device can suspect contact with a positive case.

\paragraphb{\vote.}
We propose to use the blockchain to provide an additional defense to the
\idspoofatk.
When a user generates \sysids,
she also requests blind-signatures for a set of codes
of the form $i \cncat n_i$, where $n_i$ is a nonce
and $i$ represents a sequential case number;
these codes are signed with a daily signing key different
than the signing key used for \sysids.
Specifically, the user generates sufficient codes to respond
to the maximum number of positive tests
that might reasonable affect her jurisdiction for a day.
Each such test is a nonce $n_i$.
She then builds a Merkle tree over these nonces,
by first hashing each one, then hashing adjacent pairs
to reach a tree root.
She blinds the tree root $R$ by calculating $R r_i^{e_t}$,
has it signed as $s = (R r_i^{e_t})^{d_t}$,
and unblinds the signature by calculating
$\signpos = r_i^{-1} s_i = R^{d_t}$.
When a device matches positive case number $i$,
instead of immediately notifying the user,
the device posts $n_i$, $\signpos$,
and the Merkle path required for verification to the blockchain,
allowing each user to determine the number of matches
corresponding to each positive case.
When a single positive case experiences an excessive number of matches,
the device may choose to warn the user that the match seems unlikely,
or may entirely ignore the match.

%% file: privacy.tex
\section{Privacy Analysis}\label{sec:privacy}
In our design, real IDs (such as phone numbers) 
are revealed to a signing server
before the server provides \ephids.
Since \ephids are blind-signed with a subliminal-channel-free signature,
each \ephid is indistinguishable, and could be associated with
any real ID that was presented on the day on which it was signed.

\textbf{Lemma 1}.
Given two signed \ephids, $\ephid_1$ and $\ephid_2$,
and two possible identities, $A$ and $B$,
the signer can gain no advantage on which \ephid is assigned to which identity;
that is, any attempt to assign $\ephid_i$ to $A$
succeeds with probability $1/2$.

\begin{proof} 
The signer generates a RSA key pair ($e$, $d$) and a public modulo $N$, and 
publishes the public exponent and modulo ($e$, $N$). 
The user generates a set of \ephrl identifiers, 
\{$\ephid_1$, $\dotsc$, $\ephid_n$\}, and a set of random 
numbers raised by the public exponent $e$ modulo $N$ 
to create blinding factors, 
\{$r_1^e \modn$, $\dotsc$, $r_n^e \modn$\}. 
She multiplies the \ephrl identifiers with the 
blinding factors to obtain the resulting value set $\mathcal{R}$.
\begin{equation}
\mathcal{R} := \{\ephid_1 * r_1^e \modn, \dotsc, \ephid_n * r_n^e \modn\}
\end{equation}
The set $\mathcal{R}$ is sent to the signer,
and the signer signs each \ephid in $\mathcal{R}$ using the private key $d$. 
Since $r_i^{ed} = r_i \modn$, the signer generates a signature set \sign
\begin{equation}
\mathcal{SD} := \{\ephid_1^d * r_1 \modn, ..., \ephid_n^d * r_n \modn\}
\end{equation}
The signer learns the signature set $\mathcal{SD}$ 
is associated with the user's real IDs. 
However, for any pair of values ($x$, $y$) in the 
signature set $\mathcal{SD}$
\begin{equation}
(\ephid_x^d * r_x \modn, \ephid_y^d * r_y \modn) \leftarrow \mathcal{SD}
\end{equation}
Since the blinding factors are randomly generated,
they could take on any value.
For example, if we consider two \ephids, $\ephid_1$ and $\ephid_2$,
blinded as follows:
$(\ephid_1 * r_1^e, \ephid_2 * r_2^e)$,
these are indistinguishable from
$(\ephid_2 * (\ephid_1 * \ephid_2^{-1} * r_1^e),
(\ephid_1 * (\ephid_2 * \ephid_1^{-1} * r_1^e))$.
In other words, for each \ephid $\ephid_i$
and each blinded \ephid $\ephid_j r_j^e$,
there exists a blinding factor for $\ephid_i$
such that the signing request for $\ephid_j$
would create a signature for $\ephid_i$.
Naturally, the security of the blind-signature scheme
requires that such blinding factors be computationally difficult to compute,
but they exist and are equiprobable blinding factors:
specifically, if a signer generates $N$ signatures on one day,
then for any given \ephid,
there are $N$ blinding possible blinding factors,
each of which corresponds to one blind-signature request.
Each such blinding factor is indistinguishably probable
to the signer.
\end{proof}

When a user provides \authtags to the signer,
she also includes the time interval and a nonce;
the signer therefore learns only the
which \ephids are used in each time interval,
and the correlation between nonces and \ephids.
First, if the signer is not compromised, there is no privacy loss,
because only the signer has the correlation between
\ephid and nonce.
Even when the signer is compromised,
a user adopting the full download method reveals
no information about her \authtags.
However, the size of download lists might be
multiple gigabytes if there are millions of participants.
The partial download method saves user resources
at the cost of some privacy in case of a compromised signer;
however, revealing only a small part of the nonce
(\eg 1~byte of a 16~byte nonce),
each user still retains a significant anonymity set.
Finally, the individual download method loses more privacy
in case of a compromised signer,
but the specific amount of privacy loss depends on the mechanism
by which the values are downloaded
(for example, when downloading over Tor~\cite{tor2020},
the privacy properties are the same as Tor's).

%% file: security.tex
\section{Security Analysis}\label{sec:security}
\paragraphb{Mitigate \dosatk.} 
\sysids mitigate the \dosatk by ensuring
that each \ephid stored by a user is an \ephid
generated for a valid user.
\sysid provides 
an \inplace verification to mitigate 
the \dosatk on the local storage in that the user
will reject an arbitrarily generated \ephid
that is not associated with a real identity.

%% file: evaluation.tex
\section{Evaluations}\label{sec:evaluation}

\subsection{\sysid on \dosatk}\label{sec:eval_dos}

\begin{figure}[t]
	\centering
	\mbox{
		\includegraphics[scale=0.5]{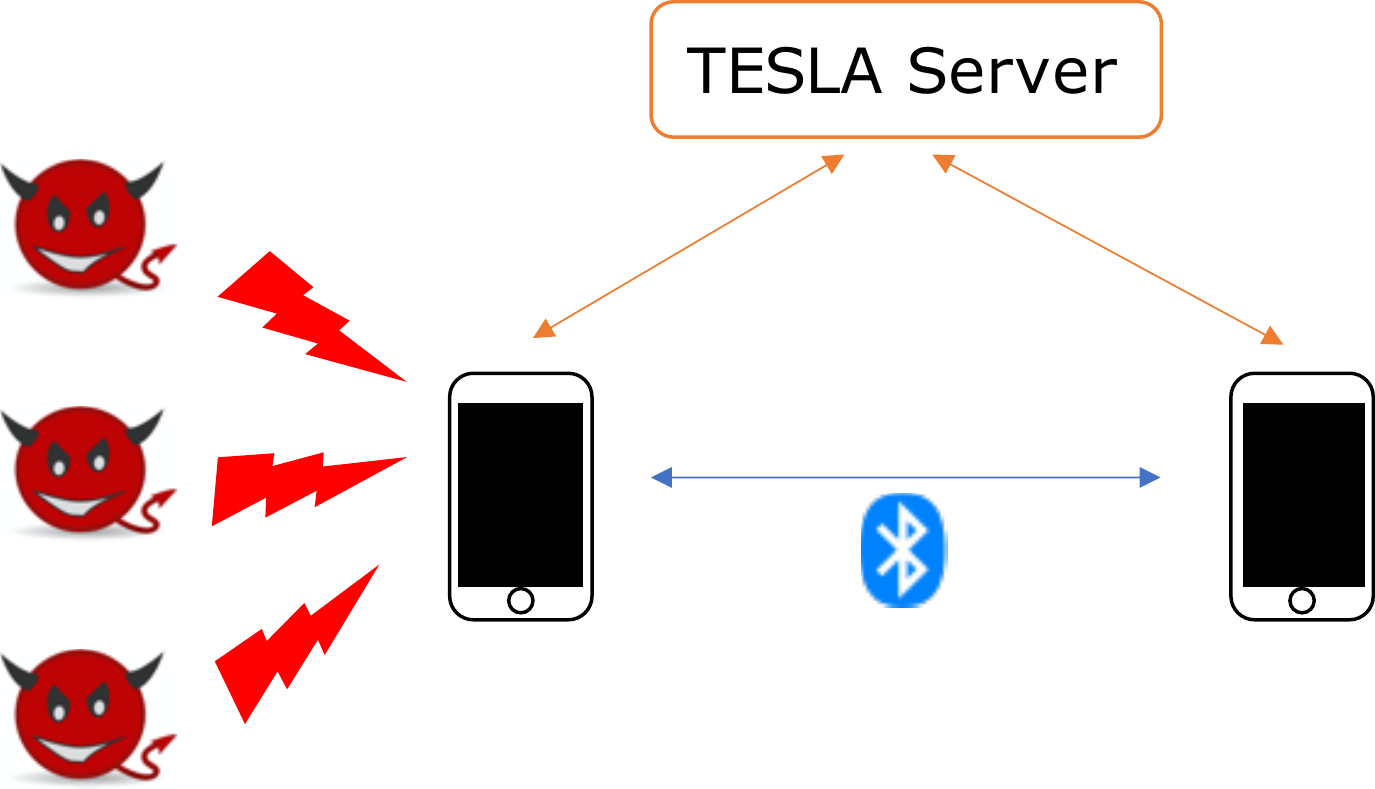}  
	}
	\caption{\dosatk Experimental Setup.} \label{fig:dos_arch}
\end{figure}

\begin{figure}[t]
	\centering
	\mbox{
		\includegraphics[width=\linewidth]{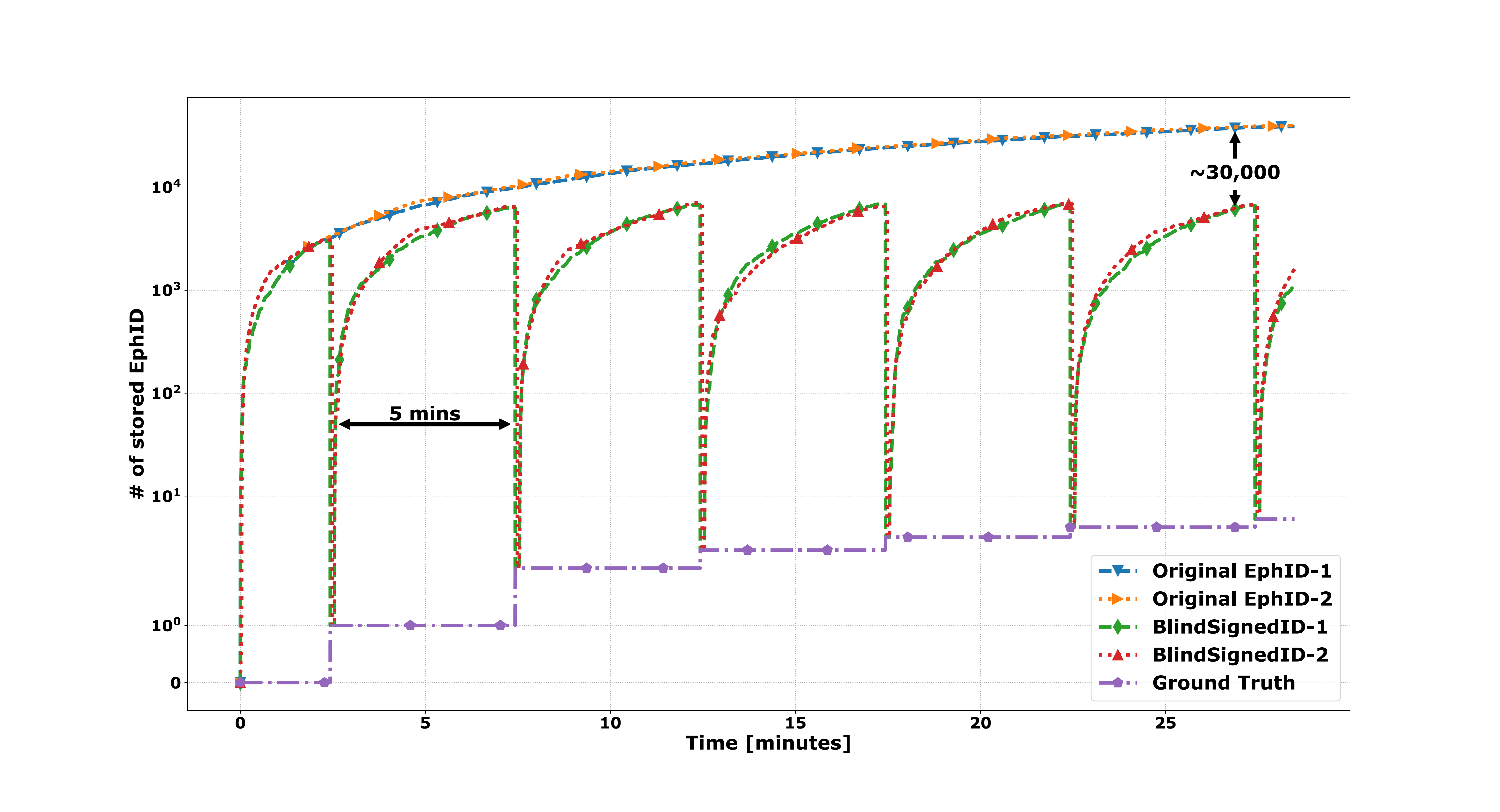}  
	}
	\caption{Stored \ephids Sent by Single Attacker.} \label{fig:dos}
\end{figure}

\paragraphb{Single attacker.}
In this section, we evaluated \sysid via sending
BLE beacons using mobile devices. Figure~\ref{fig:dos_arch}
shows our experimental setup.
We implemented a release key UDP server as the TESLA 
server, and each key was released every 5~mins. Moreover,
we built an Android application to advertise \sysid
and scan BLE beacons on two victim's phones placed
at a distance of around 20~cm. For the single attacker, 
we generated a set of 70,000 random \ephids sent 
by a single Raspberry Pi 3 Model B with the
distance 1.5~m. We used the first 13-byte of
HMAC-SHA1 as \authtag. Each random BLE beacon was broadcasted
and the rotating time was 20~ms. The receiver stored each \ephid
with the time of first receipt, duration, RSSI and \authtag
temporarily.
Whenever a key release is scheduled,
the receiver fetches the key over UDP,
checks the key,
and for each record, attempts to verify the record.
When a record is verified, the receiver puts the
time of first receipt, duration of contact, and RSSI in local storage,
consuming 38~bytes per \ephid. 
Figure~\ref{fig:dos} shows our results.
Our \sysid rejected all \dosatk periodically 
created by the attacking set, and
prevented nearly 30,000~fake \ephids stored in 30~mins. 
Consequently, the
original \ephid designs are vulnerable to \dosatk, and \sysid
can mitigate such attacks \inplace. 
In an actual attack, the attacker
can further modify BlueZ stack
to transmit BLE beacons
as frequently as possible, so \dosatk are even more severe under
such circumstances.

\begin{figure}[t]
	\begin{subfigure}[t]{\linewidth}
		\centering
		\includegraphics[width=\linewidth]{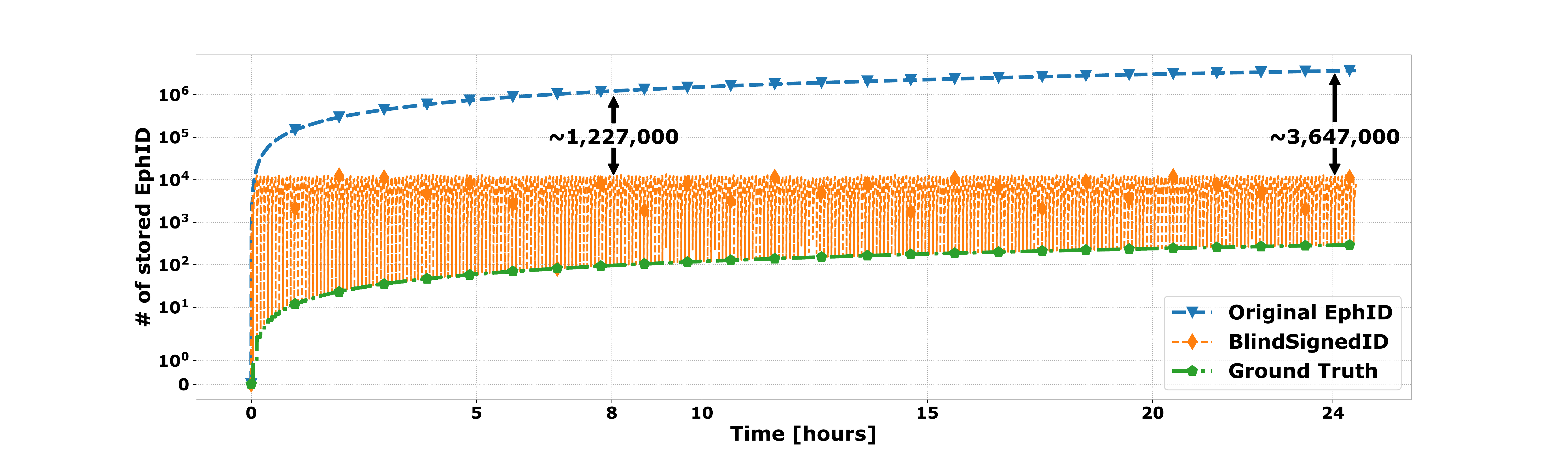}
		\caption{2 Attackers.}\label{fig:mul_dos:a}
	\end{subfigure}
	\begin{subfigure}[t]{\linewidth}
		\includegraphics[width=\linewidth]{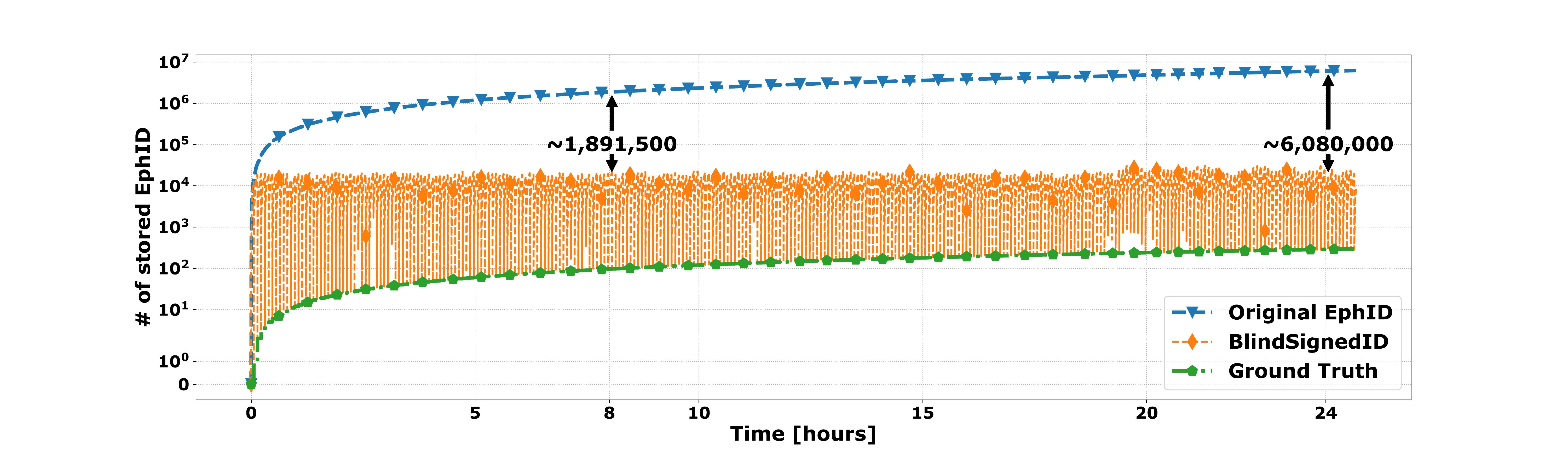}
		\caption{4 Attackers.}\label{fig:mul_dos:b}
	\end{subfigure}
	\caption{Stored \ephids Sent by Multiple Attackers.} \label{fig:mul_dos}
\end{figure}

\paragraphb{Multiple attackers.}
In addition, we evaluated \sysid with multiple attackers
to somewhat mitigate the long interval between attacker \ephids.
We generated 4 sets of 4,320,000 random \ephids, which
were used by 2 and 4 Raspberry Pis.
Figure~\ref{fig:mul_dos} shows that 2 and 4 attackers
using off-the-shelf BlueZ broadcast intervals
can cause approximately 1,227,000 and 1,891,000 stored \ephids
within 8~hrs, representing 645~MBytes and 1.076~GBytes
across 14 8-hour days, respectively. Due to networking and processing delays,
receivers do not filter all of the invalid \ephids every 5~minutes,
but the system still trends towards significant reductions in
storage requirements.

\paragraphb{Crowded environment.}
We performed 3~experiments where over 14~people were originized
in grid positions with a distance of 0.5~m, 1~m, and 1.5~m
in a middle-size room, respectively.
(These experiments were conducted in compliance with
local social distancing requirements.)
We used 6 Raspberry Pis as the
attackers to send random \ephids near the same power outlet. Each
scenario was conducted for 1--2~mins. Table~\ref{tab:large_scale}
reported the numbers of received \ephids and 
real \ephids verified by \sysid.
As distance increases, devices receive fewer \ephids
due to signal strength losses over longer distances.
Also, our \sysid continues to effectively reduce storage consumption,
removing over 90\% of the identities produced by \dosatk.

\begin{table}[t]
	\resizebox{1.01\columnwidth}{!}{%
	\begin{tabular}{c||c||c||c}
		\hline
		\begin{tabular}[c]{@{}c@{}}Scenario Distance\end{tabular} & \begin{tabular}[c]{@{}c@{}}Original \ephids\end{tabular} & \begin{tabular}[c]{@{}c@{}}\sysid\end{tabular} & \begin{tabular}[c]{@{}c@{}}Reduced Rate\end{tabular}\\ \hline \hline
		0.5~m & 12,122 & 181 & 96.72\% \\ \hline
		1.0~m & 6,526 & 76 & 97.87\% \\ \hline
		1.5~m & 2,098 & 28 & 93.59\% \\ \hline
	\end{tabular}
	} \caption{\dosatk on crowded indoor environment.}\label{tab:large_scale}
\end{table}

%% file: related.tex
\section{Related Works}\label{sec:related}
This section reviews several papers that
inspire \sysid.

\paragraphb{Digital Contact Tracing.}
\sysid mostly follows the current 
design of \dpppt~\cite{troncoso2020decentralized}, 
where a secret day seed is used to generate a 
set of \ephrl identifiers. Most current projects
follow the similar designs. 
However, current designs suffer from \dosatk. 
Although \sysid is based on \dpppt, 
it can be applied to \emph{a wide range of projects} based on 
\ephrl IDs, such as~\cite{google2020apple,mit2020pact,
pepppt2020,tcn2020,bay2020bluetrace,
chan2020pact,troncoso2020decentralized}. Moreover, 
the user registration of BlueTrace~\cite{bay2020bluetrace} 
utilizes the phone number to acquire a unique, 
randomised user ID from the backend server.
Like BlueTrace, \sysid 
requires the real IDs (\eg phone numbers), 
but the BlueTrace server can learn which 
particular phone number is linked to which user ID. 
In other words, the signer can
violate the user's privacy;
in \sysid, by contrast, the signer learns 
no information that can associate an \ephrl identifier
with a real ID,
other than the day on which the \ephid was signed
and the set of real IDs that authenticated on that day.
The ROBERT protocol~\cite{castelluccia2020robert}, 
a framework related to PEPP-PT~\cite{pepppt2020}, 
suggests that the proof-of-work system can be used 
during application registration to prevent 
from automatic registrations, but it does not 
solve the \dosatk \inplace on local storage,
and the wide range of computation power available to different devices
makes it difficult to simultaneously allow older mobile phone users
to create identities
while also preventing well-resourced attackers from creating
thousands of identities.

\paragraphb{Replay and Relay Attacks.}
\cite{pietrzak2020delayed,vaudenay2020analysis} seek to 
prevent replay and relay attacks on \dpppt. Such attacks 
can be performed by broadcasting \ephrl IDs 
generated by published secret seeds. Since these 
published seeds are positive cases, other users 
receive false alerts due to such attacks. 
\cite{vaudenay2020analysis} propose an interactive 
protocol to prevent such attacks, but it might 
not be efficient for digital contact tracing. 
\cite{pietrzak2020delayed} further present 
Delayed Authentication, which is a non-interactive 
method to prevent (long-term) replay and relay attacks.
They tie a BLE broadcast beacon
to time and location information
through a key.
The receiver verifies potential positives
with the backend server.
BlueTrace~\cite{bay2020bluetrace} utilizes 
encrypted User IDs with timestamps as 
TempID for beaconing, thus mitigating long-term replay 
attacks (but not relay attacks) by validating the 
timestamps of each TempID after the user uploads 
all records. In summary, these proposals prevent
long-term replay attacks
and potentially relay attacks by including time 
information and keys in BLE beacons. However, 
the current digital contact tracing systems 
still suffer from \dosatk.

%% file: conclusion.tex
\section{Conclusion}\label{sec:conclusion}
In this paper, we present \sysids, 
which can mitigate \dosatk
on the current digital contact tracing designs.
In particular, we propose \sysids that are \emph{verifiable}
without compromising the user's privacy to mitigate \dosatk. 
Anonymous identifiers do cause the mass surveillance 
difficult to be conducted. However, using 
anonymous identifiers \emph{without verifications} 
introduces security issues in the current designs.
Finally, we evaluated our \sysids with example
\dosatk and showd \sysids can reduce \dosatk.

%% file: acknowledgment.tex
\section*{Acknowledgment}\label{sec:ack}
This work is supported by NSF under
CNS-17313
and R.O.C. (Taiwan) Ministry of Education
for MOE 108 Government 
Scholarship to Study Abroad.
We thank Professor Hsu-Chun Hsiao for supporting
experiments at National Taiwan University
and offering valuable feedback.